\documentclass{vgtc}   

\usepackage{times}                     
\usepackage{xcolor}

\newcommand{\am}[1]{\textcolor{black}{#1}}

\usepackage{tabu}                      
\usepackage{booktabs}                  
\usepackage{lipsum}                    
\usepackage{mwe}                       
\usepackage{enumitem}
\usepackage{cite}
\usepackage[
    pagebackref,
    bookmarks,
    colorlinks=false,
    linkcolor={black},
    citecolor={black},
    urlcolor={black}]{hyperref}
\usepackage{flushend}

\onlineid{0}

\vgtccategory{Research}

\vgtcpapertype{Research}

\title{Changing the Paradigm from Dynamic Queries to\texorpdfstring{\\}{}LLM-generated SQL Queries with Human Intervention}

\author{Assor Ambre\thanks{e-mail: first-name.last-name@inria.fr. }\\ %
        \scriptsize Inria %
\and Hyeon Jeon\thanks{e-mail: hj@hcil.snu.ac.kr}\\ %
    \scriptsize SNU %
\and Sungbok Shin$^*$\\ 
     \scriptsize Inria %
\and Jean-Daniel Fekete$^*$\\ 
    \scriptsize Inria} %

\abstract{%
We propose leveraging Large Language Models (LLMs) as an interaction layer for medical visualization systems. In domains like healthcare, where users must navigate high-dimensional, coded, and heterogeneous datasets, LLM-generated queries enable \am{expert medical} users to express complex analytical intents in natural language. These intents are then translated into editable and executable queries, replacing the dynamic query interfaces used by traditional visualization systems built around sliders, check boxes, and dropdowns. 
This interaction model reduces visual clutter and eliminates the need for users to memorize field names or system codes, supporting fluid exploration, with the drawback of not exposing all the filtering criteria.
\am{We also reintroduce dynamic queries on demand to better support interactive exploration.}
We posit that \am{medical users} are trained to know the possible filtering options but challenged to remember the details of the attribute names and code values.
We demonstrate this paradigm in ParcoursVis, our scalable EventFlow-inspired patient care pathway visualization system powered by the French National Health Data System, one of the largest health data repositories in the world.
}


\teaser{
  \centering
  \includegraphics[width=0.85\linewidth]{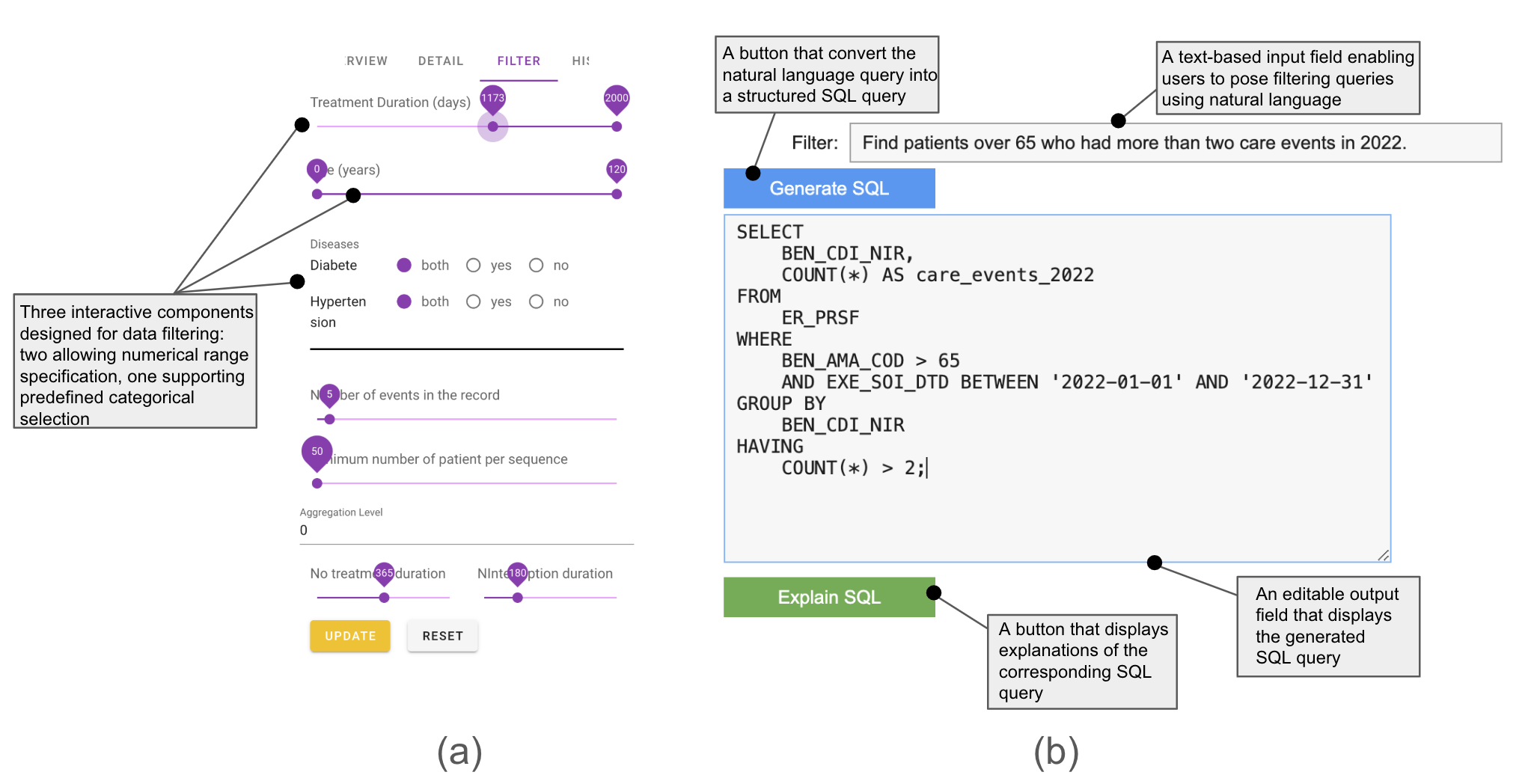}
  \caption{(a): Dynamic query widgets from \am{the current version of the} ParcoursVis aggregated EHR sequence visualization system \am{(see the Filter tab on the right panel at \href{https://parcoursvis.lisn.upsaclay.fr/}{parcoursvis.lisn.upsaclay.fr})}. The panel has limited space and cannot show the large number of available attributes and value codes for filtering patients. (b): \am{Prototype of our vision} using an LLM to translate a natural language query into an SQL query featuring an editable query box and a natural language paired-explanation.
  }\label{fig:teaser}
}

\graphicspath{} 

\usepackage{mathptmx}                  

\begin{document}



\maketitle
\section{Problem Statement} 
Large-scale Electronic Health Record (EHR) databases are being developed to inform policy-making and facilitate disease analysis~\cite{wang2022ehr}.
For \am{medical users, such as doctors, nurses, and medical administrative managers}, EHRs deliver beneficial context, such as demographics, clinical history, medications, diagnostic results, and coded data linked to external dictionaries. 
However, EHR data also presents the challenge of containing hundreds of patient attributes. 
For example, the French National Health Database (SNDS)~\cite{SNDS}, managing the EHRs of the French population,
contains over 3,000 variables related to patient care, organized in tables filled with numbers, text, and codes~\cite{SNDS2}.

Traditional visualization systems rely on the direct manipulation paradigm for interaction~\cite{shneiderman1983direct}, where users directly interact with the visualized objects of interest \am{or indirectly with dynamic query widgets~\cite{shneiderman1994dynamic}}. In EHR visualization systems such as EventFlow~\cite{Monroe:2013:TVCG} and our system ParcoursVis~\cite{ParcoursVis}, visible objects include steps in treatment pathways and event types that users can select and filter out for exploration.
Additionally, they display dynamic query widgets to filter on other attributes (\autoref{fig:teaser} (a)). 
\am{However, dynamic query user interfaces can only accommodate a limited number of widgets. }
The main challenge is to effectively deliver the specific patient information that doctors are looking for~\cite{ceneda19guidance_star}, providing sufficient filtering capabilities to support in-depth data exploration.
Given the large number of variables to manage, dynamic query interfaces may not offer the most effective solution.

\section{\am{Human-Guided LLM-generated SQL Queries}} 
To address the challenge of identifying and presenting relevant variables to \am{medical users, we envision} design choices that stem from the practical difficulties of delivering patient information in a way that aligns with their needs. Specifically, we identify two key requirements: \textit{usability} and \textit{transparency}.
Indeed, many medical users have a working knowledge of coding and database querying. 
However, navigating datasets remains difficult, particularly when variables are encoded in cumbersome ways, such as with alphanumeric codes or non-descriptive identifiers. LLMs can generate SQL queries~\cite{mohammadjafari2024natural, lee2025medorchestra, 10478355, kim2024phenoflow, li2023llmsql}, but to be effective and provide agency to doctors~\cite{ceneda2023heuristic,shneiderman2022human}, they should have these two properties:

\begin{itemize}[noitemsep,topsep=0.01cm, leftmargin=0.3cm]
    \item \textbf{Editable, scriptable queries:} The system \am{should} expose the generated SQL and allow users to edit it.
    This promotes user agency, allowing them to inspect, understand, and own the variable-selection and filtering logic~\cite{shneiderman07humanresp}.
    
    \item \textbf{Natural-language paired explanations:} Each generated query \am{should} be accompanied by a plain-language description of its effects.
    This enhances transparency~\cite{huang2025trustllm}, facilitates learning about the database structure (e.g., codes, fields), and helps detect LLM mismatches or hallucinations~\cite{huang25hallucination}.

\end{itemize}

To that end, we suggest a two-stage process to support doctors in querying complex attributes effectively on large EHR databases. 

\noindent\textbf{Step 1.} 
The first stage is the query translation phase (NL $\Rightarrow$ SQL), where the doctors' requests are translated into one or more SQL query versions. 
The translation phase is done by the LLM. The benefits are that the doctors can learn, check, edit, save the queries for later reuse, and share them with colleagues; they are deterministic, whereas the translation process with LLMs is not. 

\noindent\textbf{Step 2.} 
Within the second stage (SQL $\Rightarrow$ Vis), the queries are executed and power the visualization, which shows various details about the features the doctors selected. 

\am{Yet, this process loses some important properties of the traditional dynamic queries~\cite{shneiderman1994dynamic}.
We discuss that issue in \autoref{sec:challenges}.}

\section{Use-Case Scenario}
\am{We illustrate our vision using ParcoursVis, our web-based system for visualizing large-scale EHR event sequences. Inspired by EventFlow~\cite{Monroe:2013:TVCG}, ParcoursVis scales to datasets 4--5 orders of magnitude larger (tens of millions of patients, billions of events) using a progressive architecture~\cite{PDABook}. It is applied to \am{explore specific treatment cohorts extracted from} the French SNDS database.} 

The SNDS data are split across multiple tables (e.g., medications, procedures, hospitalizations) and use opaque alphanumeric codes, making direct querying difficult, even for users familiar with the schema. Retrieving and interpreting these codes often requires external dictionaries from various sources, which can be complex or outdated. For example, the \texttt{ER\_PRSF} table~\cite{SNDS} contains records of reimbursed healthcare acts. In our subset, each care episode includes attributes such as the beneficiary’s age at care (\texttt{BEN\_AMA\_COD}), encrypted social-security ID (\texttt{BEN\_CDI\_NIR}), year-month and exact date of death if applicable (\texttt{BEN\_DCD\_AME}, \texttt{BEN\_DCD\_DTE}), year of birth (\texttt{BEN\_NAI\_ANN}), municipality and department of residence (\texttt{BEN\_RES\_COM}, \texttt{BEN\_RES\_DPT}), and care start and end dates (\texttt{EXE\_SOI\_DTD}, \texttt{EXE\_SOI\_DTF}).

\begin{figure}\centering
    \includegraphics[width=0.8\linewidth]{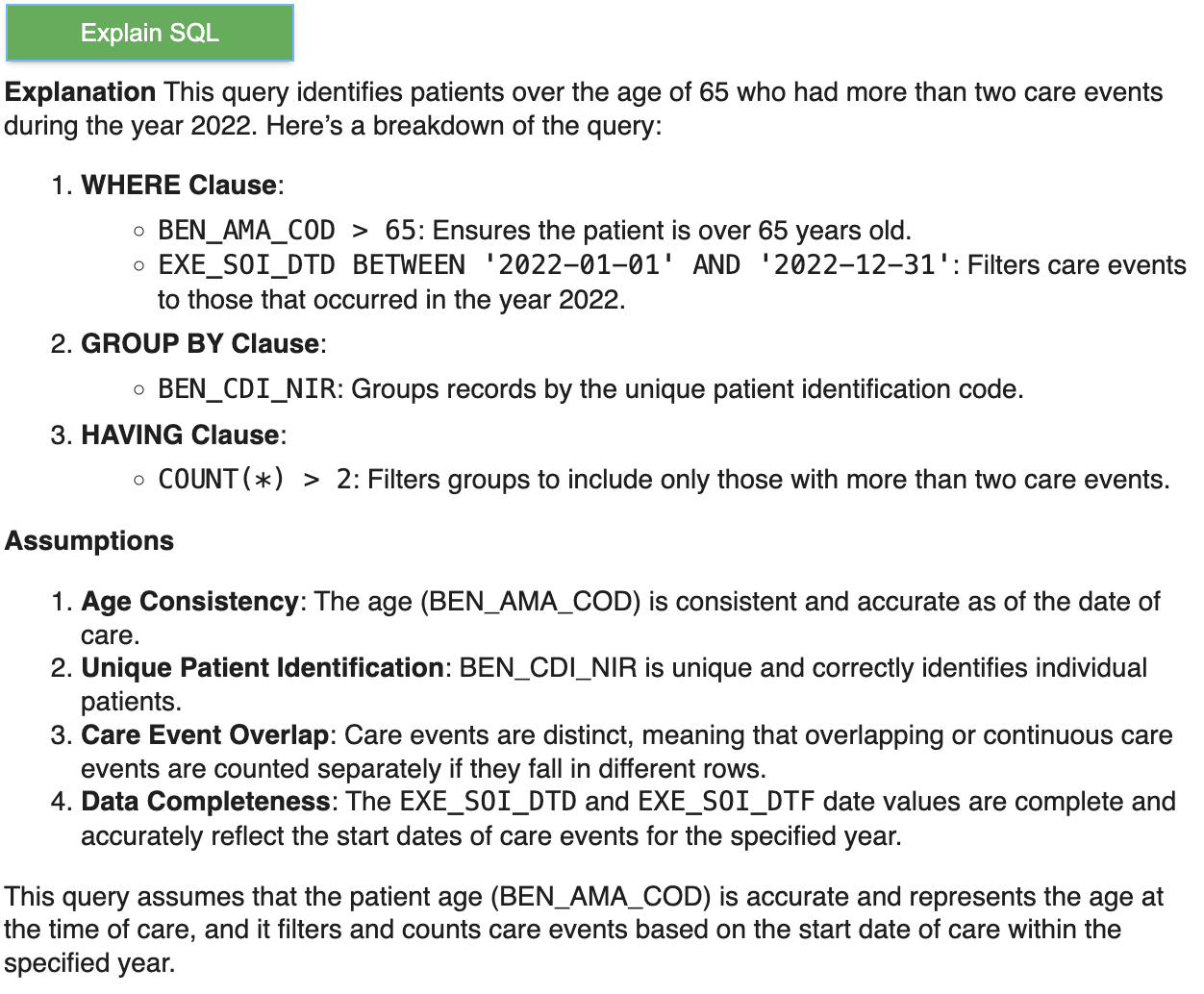}
    \caption{Example of explanations provided by our natural language to SQL LLM query translator.}\label{fig:expl}
    \vspace{-0.4cm}
\end{figure}

ParcoursVis features a \textbf{Filter tab}~(\autoref{fig:teaser} (a)) which lets users filter views by patient attributes (e.g., comorbidities, age) and by various attributes (e.g., treatment duration). This could, in principle, extend to all the available attributes from the database, but the limited screen real estate cannot present all the dynamic query widgets to enable extensive filtering of care pathways. \am{Instead, we explored the feasibility of using a pre-trained Mistral LLM~\cite{Mistral} to translate natural language queries into SQL.} It is open-source, downloadable, and licensed under Apache 2.0, so it can be run on any machine without restrictions. We provided it the \texttt{ER\_PRSF} schema and prompted it with \textit{``You are helping write SQL queries for a dataset from the SNDS health database. The data is stored in a single flat table called ``ER\_PRSF''}, asking for the following format: \textit{1.\ SQL query in a code block; 2.\ Short explanation; 3.\ Any assumptions you made.} 
\textbf{Step 1.}'s output is shown in~\autoref{fig:teaser} (b), and~\autoref{fig:expl}. \am{This step was part of an initial investigation to determine whether LLM-based query generation could be viable in our context (it is not integrated into the final system yet).}

\section{\am{Challenges Addressing the Benefits and Limitations of using LLMs in Healthcare}}\label{sec:challenges}

\noindent\textbf{Standardization of datasets. } 
Processing complex health-related datasets such as SNDS presents several challenges. 
Besides navigating multiple tables, a model must detect and resolve duplicates, distinguish them when necessary, determine the shortest join paths, and suggest alternative queries. 
We think that many of these issues are caused by the lack of standardization, and thus advocate for standardizing \am{schemas}.
Below, we introduce two examples. 

\begin{itemize}[noitemsep,topsep=0.01cm, leftmargin=0.3cm]
\item \textbf{Adopt the OMOP \am{data model}.} The ongoing effort to align health data with the OMOP model~\cite{OMOP} promotes interoperability across clinical and medico-administrative databases. Conforming SNDS and other health datasets to OMOP would supply consistent input, allowing \am{fine-tuned} models to train on OMOP data.
\item \textbf{Flatten complex datasets.} 
\am{When applications do not require the richness of the full SNDS or equivalent, we propose simplifying medical data by flattening it into smaller single-table formats to reduce ambiguity, simplify analysis, and simplify LLM-based SQL generation. 
Different analytical needs (e.g., patient, disease, or treatment-based) may benefit from tailored flat tables, suggesting a use-case-specific flattening strategy.}
\end{itemize}

\smallskip

\noindent\textbf{Back to dynamic query interfaces. }\am{To revive key properties of dynamic queries, in particular \emph{rapid, incremental, and reversible control; selection by pointing, not typing; and immediate, continuous feedback}, we propose extracting SQL subtrees, based on prior queries or user requests, to generate dynamic query widgets. Such a mechanism, already explored in~\cite{pister2023combinet}, supports customizable, context-aware queries built from SQL fragments in the original tree. This implementation of the interface logic could be simplified by driving widget generation by an LLM~\cite{masson2024directGPT} (versus being manually coded).}

\section{Conclusion}
We outline our position in proposing an approach to replace pure dynamic query interfaces with LLM-generated, editable, and explained queries.
It would promote user agency, allowing users to inspect and understand the variable-selection and filtering logic, and mitigate the unavoidable hallucinations.
Our approach preserves the possibility of using regular dynamic queries for an interactively selected subset of queries, but provides the scalability needed to deal with real-world EHR databases in exploratory settings.

\section*{Acknowledgements}
We thank our reviewers for their valuable insights. This work was supported in part by a grant from the Health Data-Hub, and from the \href{https://www.inria.fr/en/urge}{URGE AP-HP/Inria project}.

\bibliographystyle{abbrv-doi-hyperref}

\bibliography{template}

\begin{thebibliography}{10}

\bibitem{ParcoursVis}
A.~Assor, M.~Sereno, and J.-D. Fekete.
\newblock \href{https://arxiv.org/abs/2508.10700}{Visualization of electronic health record sequences at scale}.
\newblock {\em arXiv preprint arXiv:2508.10700}, 2025.

\bibitem{10478355}
R.~C. Basole and T.~Major.
\newblock \href{https://doi.org/10.1109/MCG.2024.3362168}{{Generative AI for Visualization: Opportunities and Challenges}}.
\newblock {\em IEEE Computer Graphics and Applications}, 44(2):55--64, 2024. \href{https://doi.org/10.1109/MCG.2024.3362168}
{doi: {{%
10\hspace{.1pt}\discretionary{.}{%
}{.}\hspace{.4pt}1109\discretionary{/}{%
}{/}MCG\hspace{.1pt}\discretionary{.}{%
}{.}\hspace{.4pt}2024\hspace{.1pt}\discretionary{.}{%
}{.}\hspace{.4pt}3362168}}}


\bibitem{ceneda2023heuristic}
D.~Ceneda, C.~Collins, M.~El-Assady, S.~Miksch, C.~Tominski, and A.~Arleo.
\newblock A heuristic approach for dual expert/end-user evaluation of guidance in visual analytics.
\newblock {\em {IEEE} Trans. Vis. Comput. Graph.}, 30(1):997--1007, 2023.

\bibitem{ceneda19guidance_star}
D.~Ceneda, T.~Gschwandtner, and S.~Miksch.
\newblock \href{https://doi.org/https://doi.org/10.1111/cgf.13730}{A review of guidance approaches in visual data analysis: A multifocal perspective}.
\newblock {\em Computer Graphics Forum}, 38(3):861--879, 2019. \href{https://doi.org/10.1111/cgf.13730}
{doi: {{%
10\hspace{.1pt}\discretionary{.}{%
}{.}\hspace{.4pt}1111\discretionary{/}{%
}{/}cgf\hspace{.1pt}\discretionary{.}{%
}{.}\hspace{.4pt}13730}}}


\bibitem{PDABook}
J.-D. Fekete, D.~Fisher, and M.~Sedlmair.
\newblock {\em Progressive Data Analysis: Roadmap and Research Agenda}.
\newblock Eurographics, Nov. 2024. \href{https://doi.org/10.2312/pda.20242707}
{doi: {{%
10\hspace{.1pt}\discretionary{.}{%
}{.}\hspace{.4pt}2312\discretionary{/}{%
}{/}pda\hspace{.1pt}\discretionary{.}{%
}{.}\hspace{.4pt}20242707}}}


\bibitem{huang25hallucination}
L.~Huang, W.~Yu, W.~Ma, W.~Zhong, Z.~Feng, H.~Wang, Q.~Chen, W.~Peng, X.~Feng, B.~Qin, and T.~Liu.
\newblock \href{https://doi.org/10.1145/3703155}{A survey on hallucination in large language models: Principles, taxonomy, challenges, and open questions}.
\newblock {\em ACM Trans. Inf. Syst.}, 43(2),  article no. 42,  55 pages, Jan. 2025. \href{https://doi.org/10.1145/3703155}
{doi: {{%
10\hspace{.1pt}\discretionary{.}{%
}{.}\hspace{.4pt}1145\discretionary{/}{%
}{/}3703155}}}


\bibitem{huang2025trustllm}
Y.~Huang, L.~Sun, H.~Wang, S.~Wu, Q.~Zhang, Y.~Li, C.~Gao, Y.~Huang, W.~Lyu, et~al.
\newblock \href{https://openreview.net/forum?id=bWUU0LwwMp}{{Position: TrustLLM: Trustworthiness in Large Language Models}}.
\newblock In {\em Forty-first International Conference on Machine Learning, ({ICML})}. OpenReview.net, 2024.

\bibitem{kim2024phenoflow}
J.~Kim, S.~Lee, H.~Jeon, K.-J. Lee, H.-J. Bae, B.~Kim, and J.~Seo.
\newblock {PhenoFlow: A Human-LLM Driven Visual Analytics System for Exploring Large and Complex Stroke Datasets}.
\newblock {\em {IEEE} Trans. Vis. Comput. Graph.}, 2024.

\bibitem{lee2025medorchestra}
S.~Lee, H.~Song, J.-c. Lee, Y.~J. Lee, B.~Lee, H.-E. Lim, D.~Kim, J.~Seo, and B.~Kim.
\newblock {MedOrchestra: A Hybrid Cloud-Local LLM Approach for Clinical Data Interpretation}.
\newblock {\em arXiv preprint arXiv:2505.23806}, 2025.

\bibitem{li2023llmsql}
J.~Li, B.~Hui, G.~Qu, J.~Yang, B.~Li, B.~Li, B.~Wang, B.~Qin, R.~Geng, N.~Huo, X.~Zhou, C.~Ma, G.~Li, K.~C. Chang, F.~Huang, R.~Cheng, and Y.~Li.
\newblock \href{http://papers.nips.cc/paper\_files/paper/2023/hash/83fc8fab1710363050bbd1d4b8cc0021-Abstract-Datasets\_and\_Benchmarks.html}{{Can {LLM} Already Serve as {A} Database Interface? {A} BIg Bench for Large-Scale Database Grounded Text-to-SQLs}}.
\newblock In {\em Advances in Neural Information Processing Systems 36: Annual Conference on Neural Information Processing Systems, (NeurIPS)}, 2023.

\bibitem{masson2024directGPT}
D.~Masson, S.~Malacria, G.~Casiez, and D.~Vogel.
\newblock \href{https://doi.org/10.1145/3613904.3642462}{Directgpt: A direct manipulation interface to interact with large language models}.
\newblock In {\em Proceedings of the ACM Conference on Human Factors in Computing Systems},  article no. 975,  16 pages. Association for Computing Machinery, New York, NY, USA, 2024. \href{https://doi.org/10.1145/3613904.3642462}
{doi: {{%
10\hspace{.1pt}\discretionary{.}{%
}{.}\hspace{.4pt}1145\discretionary{/}{%
}{/}3613904\hspace{.1pt}\discretionary{.}{%
}{.}\hspace{.4pt}3642462}}}


\bibitem{Mistral}
Mistral-small-3.1-24b-instruct-2503.
\newblock \url{https://huggingface.co/mistralai/Mistral-Small-3.1-24B-Instruct-2503}.
\newblock Accessed: 2025-06-30.

\bibitem{mohammadjafari2024natural}
A.~Mohammadjafari, A.~S. Maida, and R.~Gottumukkala.
\newblock \href{https://arxiv.org/abs/2410.01066}{{From natural language to SQL: Review of LLM-based text-to-SQL systems}}.
\newblock {\em arXiv preprint arXiv:2410.01066}, 2024.

\bibitem{Monroe:2013:TVCG}
M.~Monroe, R.~Lan, H.~Lee, C.~Plaisant, and B.~Shneiderman.
\newblock \href{https://doi.org/10.1109/TVCG.2013.200}{{Temporal Event Sequence Simplification}}.
\newblock {\em {IEEE} Trans. Vis. Comput. Graph.}, 19(12):2227--2236, 2013. \href{https://doi.org/10.1109/TVCG.2013.200}
{doi: {{%
10\hspace{.1pt}\discretionary{.}{%
}{.}\hspace{.4pt}1109\discretionary{/}{%
}{/}TVCG\hspace{.1pt}\discretionary{.}{%
}{.}\hspace{.4pt}2013\hspace{.1pt}\discretionary{.}{%
}{.}\hspace{.4pt}200}}}


\bibitem{OMOP}
{Standardized Data: The OMOP Common Data Model}.
\newblock \url{https://www.ohdsi.org/data-standardization/}.
\newblock Accessed: 2025-06-30.

\bibitem{pister2023combinet}
A.~Pister, C.~Prieur, and J.-D. Fekete.
\newblock \href{https://doi.org/10.1111/CGF.14731}{{ComBiNet: Visual Query and Comparison of Bipartite Multivariate Dynamic Social Networks}}.
\newblock {\em Computer Graphics Forum}, 42(1):290--304, 2023. \href{https://doi.org/10.1111/CGF.14731}
{doi: {{%
10\hspace{.1pt}\discretionary{.}{%
}{.}\hspace{.4pt}1111\discretionary{/}{%
}{/}CGF\hspace{.1pt}\discretionary{.}{%
}{.}\hspace{.4pt}14731}}}


\bibitem{shneiderman1983direct}
B.~Shneiderman.
\newblock \href{https://doi.org/10.1109/MC.1983.1654471}{{Direct Manipulation: A Step Beyond Programming Languages}}.
\newblock {\em Computer}, 16(8):57–69,  13 pages, Aug. 1983. \href{https://doi.org/10.1109/MC.1983.1654471}
{doi: {{%
10\hspace{.1pt}\discretionary{.}{%
}{.}\hspace{.4pt}1109\discretionary{/}{%
}{/}MC\hspace{.1pt}\discretionary{.}{%
}{.}\hspace{.4pt}1983\hspace{.1pt}\discretionary{.}{%
}{.}\hspace{.4pt}1654471}}}


\bibitem{shneiderman1994dynamic}
B.~Shneiderman.
\newblock \href{https://doi.org/10.1109/52.329404}{Dynamic queries for visual information seeking}.
\newblock {\em IEEE Software}, 11(6):70--77, 1994. \href{https://doi.org/10.1109/52.329404}
{doi: {{%
10\hspace{.1pt}\discretionary{.}{%
}{.}\hspace{.4pt}1109\discretionary{/}{%
}{/}52\hspace{.1pt}\discretionary{.}{%
}{.}\hspace{.4pt}329404}}}


\bibitem{shneiderman07humanresp}
B.~Shneiderman.
\newblock \href{https://doi.org/10.1109/MIS.2007.32}{Human responsibility for autonomous agents}.
\newblock {\em IEEE Intelligent Systems}, 22(2):60--61, March 2007. \href{https://doi.org/10.1109/MIS.2007.32}
{doi: {{%
10\hspace{.1pt}\discretionary{.}{%
}{.}\hspace{.4pt}1109\discretionary{/}{%
}{/}MIS\hspace{.1pt}\discretionary{.}{%
}{.}\hspace{.4pt}2007\hspace{.1pt}\discretionary{.}{%
}{.}\hspace{.4pt}32}}}


\bibitem{shneiderman2022human}
B.~Shneiderman.
\newblock {\em Human-centered AI}.
\newblock Oxford University Press, 2022.

\bibitem{SNDS2}
{Documentation collaborative du SNDS}.
\newblock \url{https://documentation-snds.health-data-hub.fr/}.
\newblock Accessed: 2025-06-30.

\bibitem{SNDS}
P.~Tuppin, J.~Rudant, P.~Constantinou, C.~Gastaldi-Ménager, A.~Rachas, L.~{de Roquefeuil}, G.~Maura, H.~Caillol, A.~Tajahmady, J.~Coste, C.~Gissot, A.~Weill, and A.~Fagot-Campagna.
\newblock \href{https://doi.org/10.1016/j.respe.2017.05.004}{{Value of a national administrative database to guide public decisions: From the système national d’information interrégimes de l’Assurance Maladie (SNIIRAM) to the système national des données de santé (SNDS) in France}}.
\newblock {\em Revue d'Épidémiologie et de Santé Publique}, 65:S149--S167, 2017.
\newblock Réseau REDSIAM. \href{https://doi.org/10.1016/j.respe.2017.05.004}
{doi: {{%
10\hspace{.1pt}\discretionary{.}{%
}{.}\hspace{.4pt}1016\discretionary{/}{%
}{/}j\hspace{.1pt}\discretionary{.}{%
}{.}\hspace{.4pt}respe\hspace{.1pt}\discretionary{.}{%
}{.}\hspace{.4pt}2017\hspace{.1pt}\discretionary{.}{%
}{.}\hspace{.4pt}05\hspace{.1pt}\discretionary{.}{%
}{.}\hspace{.4pt}004}}}


\bibitem{wang2022ehr}
Q.~Wang and R.~S. Laramee.
\newblock \href{https://doi.org/10.1111/CGF.14424}{{EHR} {STAR:} the state-of-the-art in interactive {EHR} visualization}.
\newblock {\em Computer Graphics Forum}, 41(1):69--105, 2022. \href{https://doi.org/10.1111/CGF.14424}
{doi: {{%
10\hspace{.1pt}\discretionary{.}{%
}{.}\hspace{.4pt}1111\discretionary{/}{%
}{/}CGF\hspace{.1pt}\discretionary{.}{%
}{.}\hspace{.4pt}14424}}}


\end{thebibliography}

\typeout{get arXiv to do 4 passes: Label(s) may have changed. Rerun}

\end{document}